\documentclass[
  ,draft            
  ]{aipproc}

\layoutstyle{6x9}

\begin{document}

\title[Spontaneous Dimensional Reduction in Short-Distance 
Quantum Gravity?]{Spontaneous Dimensional Reduction\\ 
in Short-Distance Quantum Gravity?}


\author{Steven Carlip}{
  address={Physics Department\\University of California at Davis\\Davis, 
CA 95616, USA}
}
 
\begin{abstract}
Several lines of evidence suggest that quantum gravity at very short
distances may behave effectively as a two-dimensional theory.  I 
summarize these hints, and offer an additional argument based on 
the strong-coupling limit of the Wheeler-DeWitt equation.  The 
resulting scenario suggests a novel approach to quantum gravity 
at the Planck scale.
\end{abstract}

\maketitle

At large scales, spacetime behaves as a smooth four-dimensional 
manifold.  At the Planck scale, on the other hand, the appropriate
description is not so clear: we have neither observational evidence
nor an established theoretical framework, and it is not even   
obvious that ``space'' and ``time'' are proper categories.

But while a complete quantum theory of gravity remains distant,  
we have a number of fragments that may offer hints.  When these 
fragments fit together---when a fundamental feature of spacetime 
appears robustly across different approaches to quantum gravity%
---we should consider the possibility that our models are telling us 
something real about Nature.  The thermodynamic behavior of black 
holes, for example, occurs so consistently, across so many different 
approaches, that it is reasonable to expect quantum gravity to
provide a statistical mechanical explanation.

Over the past few years, evidence has accumulated that spacetime
near the Planck scale is effectively two-dimensional.  No single  
indication of this behavior is in itself very convincing, but taken 
together, they may point toward a promising direction for further 
investigation.  Here, I will summarize these hints, and provide a 
new piece of evidence in the form of a strong-coupling approximation 
to the Wheeler-DeWitt equation.

\section{Is small-scale quantum gravity two-dimensional?}

Evidence for ``spontaneous dimensional reduction'' at short distances 
comes from a variety of different approaches to quantum gravity.  
Among these are the following:\\

\noindent{\bf Causal Dynamical Triangulations}\\[-1.5ex]

\begin{figure}
\includegraphics[width=4.5in]{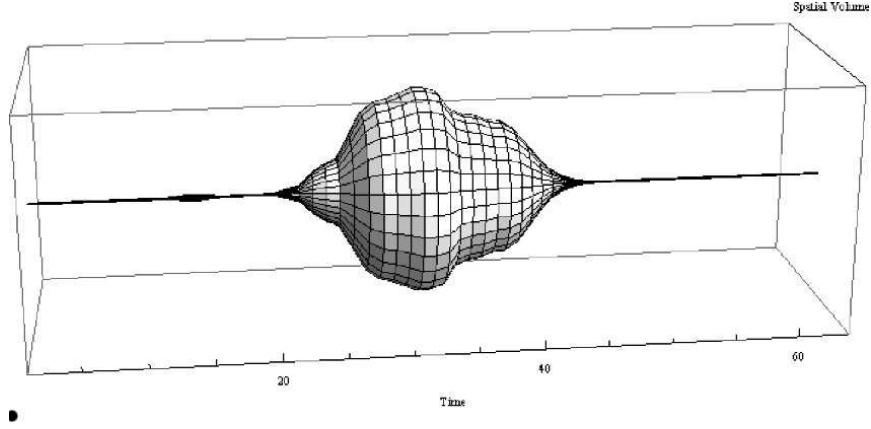} 
\caption{A contribution to the path integral in causal dynamical 
triangulations}
\source{R.\ Kommu, U.C.\ Davis}
 \label{CDT}
\end{figure}

The ``causal dynamical triangulation'' program \cite{AJL,AJL2,AJL3} 
is a discrete approximation to the gravitational path integral, in 
which the spacetimes contributing to the sum over histories are 
approximated by locally flat simplicial manifolds.  The idea of  a 
simplicial approximation dates back to Regge's work in the 1960s 
\cite{Regge}, and the suggestion of using Monte Carlo calculations
was made as early as 1981 \cite{Williams}.  Until fairly recently, 
though, such efforts failed, yielding only a ``crumpled'' phase with a 
very high Hausdorff dimension and a two-dimensional ``branched 
polymer'' phase \cite{Loll}.  The crucial new ingredient introduced 
by Ambj{\o}rn et al.\ is a definite causal structure, in the form 
of a fixed time-slicing.  The resulting path integral appears to lead 
to four-dimensional spacetimes, with contributions of the form shown 
in figure \ref{CDT} \cite{Kommu}; moreover, the computed cosmological 
scale factor has the correct semiclassical behavior \cite{AJL3,AJL4}.

A key question for any such discrete approach is whether it genuinely
reproduces the four-dimensional structure we observe at large distances.  
This is a subtle issue, requiring a definition of dimension for a discrete 
structure that may be very non-manifold-like at short distances.   One 
natural choice is the spectral dimension \cite{spec}, the dimension as 
seen by a diffusion process or a random walker.  

Diffusion from an initial position $x$ to a final position $x'$ in time
$s$ may be described by a heat kernel $K(x,x',s)$, satisfying
\begin{equation}
\left(\frac{\partial\ }{\partial s} - \Delta_x\right)K(x,x';s) =0,
\qquad \hbox{with \quad $K(x,x',0) = \delta(x-x')$} .
\label{Ca1}
\end{equation}
On a manifold of dimension $d_S$, the heat kernel generically behaves as
\begin{equation}
K(x,x';s) \sim (4\pi s)^{-d_S/2} e^{-\sigma(x,x')/2s}
    \left( 1 + \mathcal{O}(s)\right)
\label{Ca2}
\end{equation}
for small $s$, where $\sigma(x,x')$ is Synge's ``world function''
\cite{Synge}, one-half of the geodesic distance between $x$ and $x'$.
In particular, the return probability $K(x,x,s)$ is
\begin{equation}
K(x,x;s) \sim (4\pi s)^{-d_S/2} .
\label{Ca3}
\end{equation}
For any space on which a diffusion process can be defined, we can then 
use equation (\ref{Ca3}) to define an effective dimension $d_S$, the 
spectral dimension.

For the causal dynamical triangulation program, the spectral dimension
is measured, to within numerical accuracy, to be $d_S=4$ at large
distances \cite{AJL3,spec}.  This is a promising sign, indicating the
recovery of four-dimensional behavior.  At short distances, though, the 
spectral dimension falls to two.  A similar behavior occurs in 
(2+1)-dimensional gravity \cite{Kommu}.  This is our first indication 
of dimensional reduction at short distances.

Now, the spectral dimension is not the unique generalization of dimension,
and one may worry about reading too much significance into this result.
Note, though, that the propagator for a scalar field may be obtained 
as a Laplace transform of the heat kernel.  The behavior of the spectral 
dimension then leads to a propagator
\begin{equation}
G(x,x') \sim \int_0^\infty \!ds\, K(x,x';s) \sim \left\{\begin{array}{cl}
   \sigma^{-2} \quad & \hbox{at large distances}\\
   \ln\sigma \quad & \hbox{at small distances.} \end{array}\right. 
\label{Ca4}
\end{equation}
The logarithmic short-distance behavior is characteristic of a
two-dimensional field theory, and strongly suggests that if one probes 
short distances with quantum fields, one will measure an effective 
dimension of two.\\

\noindent{\bf Renormalization Group Analysis}\\[-1.5ex]

General relativity is, of course, nonrenormalizable.  Nevertheless, a
renormalization group analysis may give us useful information about
quantum gravity.  In particular, Weinberg has suggested that the theory 
may be ``asymptotically safe'' \cite{Weinberg}.  

Consider the full effective action for metric gravity, containing 
an infinite number of higher-derivative terms with an infinite number
of coupling constants.  Under the renormalization group flow, some
of these constants may blow up, indicating that the effective action 
description has broken down and new physics is needed.  It could 
be, however, that the coupling constants remain finite and flow to an 
ultraviolet fixed point.  In that case, the theory would continue to 
make sense down to arbitrarily short distances.  If, in addition, the 
critical surface---the space of such UV fixed points---were finite 
dimensional, the coupling constants would be determined by a finite 
number of parameters: not quite renormalizability, but almost as good.

We do not yet know whether quantum general relativity exhibits such 
behavior.  But renormalization group flows of a variety of truncated 
actions offer evidence of a UV fixed point \cite{Reuter, Litim,Nieder}.  
For the present investigation, the key feature of these results is that 
operators obtain large anomalous dimensions at the fixed point---precisely 
the dimensions that characterize a two-dimensional field theory 
\cite{Reuter}.  Moreover, a computation of the spectral dimension 
near the putative fixed point again yields $d_S=2$ \cite{Reuter2}.

There is, in fact, a fairly general argument that if quantum gravity
is asymptotically safe, it must behave like a two-dimensional theory
at the UV fixed point \cite{Nieder}.  Consider the dimensionless
coupling constant $g_N(\mu) = G_N\mu^{d-2}$, where $G_N$ is
Newton's constant.  Under renormalization group flow,
\begin{equation}
\mu\frac{\partial g_N}{\partial\mu} 
   = [d-2+\eta_N(g_N,\dots)] g_N  ,
\label{Ca5}
\end{equation}
where the anomalous dimension $\eta_N$ depends on both $g_N$ and 
any other dimensionless coupling constants in the theory.  For a 
non-Gaussian fixed point $g_N^*$ to occur,\footnote{``Non-Gaussian''
simply means ``not free field,'' i.e., $0<g_N^*<\infty$.} the right-hand
side of (\ref{Ca5}) must vanish, that is, $\eta_N(g_N^*,\dots) = 2-d$.

But the propagator of a field with anomalous dimension $\eta_N$  has
a momentum dependence of the form $(p^2)^{-1 + \eta_N/2}$.  For
$\eta_N = 2-d$, this is $p^{-d}$, and the corresponding position space 
propagator depends logarithmically on distance.  Such behavior is, again, 
the characteristic of a two-dimensional field.  While the argument I have
given applies to the graviton propagator, a generalization to arbitrary
fields is straightforward \cite{Nieder}.\\

\noindent{\bf Loop quantum gravity}\\[-1.5ex]

Our next indication of short-distance dimensional reduction comes from the 
area spectrum of loop quantum gravity \cite{Modesto}.  This spectrum 
is labeled by half-integers $j$: $A_j \sim \ell_p^2\sqrt{j(j+1)}$, where 
$\ell_p$ is the Planck length.  Defining $\ell_j = \sqrt{j}\ell_p$, we can 
rewrite the spectrum as
\begin{equation}
A_j \sim \sqrt{\ell_j^2(\ell_j^2 + \ell_p^2)} \sim \left\{\begin{array}{cl}
   \ell_j^2 \quad & \hbox{for large areas}\\
   \ell_p\ell_j \quad & \hbox{for small areas.} \end{array}\right. 
\label{Ca6}
\end{equation}
Like the propagator (\ref{Ca4}), this spectrum undergoes a change
in scaling at small distances.  Modesto argues that this behavior 
determines the scaling of an effective metric, and uses the this scaling 
to compute a spectral dimension.  The result is again an effective 
dimension that decreases from four at large scales to two at small scales.\\

\noindent{\bf High temperature strings}\\[-1.5ex]

Yet another piece of evidence comes from the high temperature behavior 
of string theory.  In 1988, Atick and Witten showed that at temperatures 
far above the Hagedorn temperature, string theory has a very peculiar
thermodynamic behavior \cite{Atick}: the free energy in a volume $V$ 
varies with temperature as
\begin{equation}
F/VT \sim T .
\label{Ca7}
\end{equation}
For a field theory in $d$ dimensions, in contrast, $F/VT\sim T^{d-1}$.
Thus, although string theory lives in 10 or 26 dimensions, at high 
temperatures it behaves in some ways as if spacetime were 
two-dimensional.\\

\noindent{\bf Anisotropic scaling models}\\[-1.5ex]

As a final indication of spontaneous dimensional reduction, we can
consider ``Ho{\v r}ava-Lifshitz gravity'' \cite{Horava}, a set of  new 
models of gravity that exhibit anisotropic scaling, that is, invariance 
under rescalings ${\bf x}\rightarrow b{\bf x},\ t\rightarrow b^3t$.
Such a scaling property clearly breaks Lorentz invariance (which may, 
however, be restored at low energies).  In fact, this symmetry breaking 
is the key to renormalizability: the field equations may contain many 
spatial derivatives, giving high inverse powers of spatial momentum 
to tame loop integrals, while keeping only second time derivatives, 
thus avoiding ghosts.

Ho{\v r}ava has calculated the spectral dimension in such models
\cite{Horava2}, and finds that $d_S=2$ at high energies.  In one
sense, this is a cautionary tale: the ``two-dimensional'' behavior 
arises from the fact that the propagators contain higher inverse 
powers of momentum, and the logarithmic dependence on distance
comes from integrals of the form $\int d^4p/p^4$ rather than from
any intrinsic two-dimensional structure.  The lesson, I believe, is 
that ``dimension'' is not such an obvious quantity in quantum gravity, 
but may have different meanings depending on how one probes the 
physics.  In particular, despite the four-dimensional origin of the
spectral dimension in these models, the results imply that quantum
fields will behave ``two-dimensionally'' at short distances.

\section{The strong-coupling approximation}

Suppose the hints of the preceding section are really telling us 
something deep about short-distance quantum gravity.  A number of 
obvious questions arise.  Most strikingly, we may ask, ``Which two 
dimensions?''  How can a theory with a four-dimensional Lorentz 
symmetry pick out two ``preferred'' directions at small scales?  To 
address this question, let us consider one more approach to physics at 
the Planck scale: the strong-coupling approximation of the 
Wheeler-DeWitt equation.

As early as 1976, Isham \cite{Isham} noted that the Wheeler-DeWitt 
equation \cite{DeWitt}
\begin{equation}
\left\{ 16\pi\ell_p^2G_{ijkl}
    \frac{\delta\ }{\delta g_{ij}} \frac{\delta\ }{\delta g_{kl}}
    - \frac{1}{16\pi\ell_p^2}\sqrt{g}\,{}^{(3)}\!R\right\}\Psi[g] = 0 
\label{Ca8}
\end{equation}
has an interesting strong-coupling limit $\ell_p\rightarrow\infty$.
In this limit, the Wheeler-DeWitt equation becomes ultralocal: spatial
derivatives appear only in the scalar curvature term, and when this 
term drops out, points effectively decouple.  As Pilati first observed 
\cite{Pilati}, this limit probes spacetime near or below the Planck 
scale; Maeda and Sakamoto have expounded this argument in more 
detail \cite{Maeda}.

The $\ell_p=\infty$ limit of the Wheeler-DeWitt equation was studied 
extensively in the 1980s \cite{HPT,Teitelboim,Pilati2,Rovelli,Husain}
and a perturbative treatment of the scalar curvature term has been 
discussed by several authors \cite{Francisco,Salopek,Kirillov}.  The 
key features are already evident in the classical version.  The 
strong-coupling approximation can also be viewed as a small $c$ 
approximation; as $\ell_p$ becomes large, light cones contract
to timelike lines, and neighboring points decouple \cite{Henneaux}.
The classical solution at each point is a Kasner space,
\begin{eqnarray}
&&ds^2 = dt^2 - t^{2p_1}dx^2 - t^{2p_2}dy^2 - t^{2p_3}dz^2 \\
\label{Ca9}
&&\qquad\hbox{\small ($-\frac{1}{3}<p_1<0<p_2<p_3,\quad
p_1+p_2+p_3 = 1 = p_1^2+p_2^2+p_3^2$).}\nonumber
\end{eqnarray}
More precisely---see, for example, \cite{Helfer}---the general solution 
is an arbitrary $\mathrm{GL}(3)$ transformation of a Kasner metric: 
in effect, a Kasner space with arbitrary, not necessary orthogonal, axes.  

For large but finite $\ell_p$, the classical solution exhibits BKL 
behavior \cite{BKL,HUR}.  At each point, the metric spends most of 
its time in a nearly Kasner form.  But the scalar curvature can grow 
abruptly, making the curvature term in (\ref{Ca8}) important and leading 
to a Mixmaster-like ``bounce'' \cite{Misner} to a new Kasner solution 
with different axes and exponents.  Neighboring points are no 
longer completely decoupled, but the Mixmaster bounces are chaotic  
\cite{Chernoff}; the geometries at nearby points quickly become
uncorrelated, with Kasner exponents occurring randomly with a known 
probability distribution \cite{Kirillov2}.

We can now return to the problem of dimensional reduction.  Consider
a timelike geodesic in Kasner space, starting at $t=t_0$ with a randomly 
chosen initial velocity.  It is not hard to show that in the direction of
decreasing $t$, the proper distance along the geodesic in the direction 
of each of the Kasner axes asymptotes to
\begin{equation}
\begin{array}{l} s_x \sim t^{p_1}\\
                                  s_y \sim 0\\
                                  s_z \sim 0 . \end{array}
\label{Ca10}
\end{equation}
The geodesic effectively explores only one spatial dimension.  In the
direction of increasing $t$, a similar, though less dramatic, phenomenon
occurs:
\begin{equation}
\begin{array}{l} s_x  \sim t\\ 
                                  s_y  \sim t^{\max(p_2,1+p_1-p_2)}\\
                                  s_z  \sim t^{p_3} . \end{array}
\label{Ca11}
\end{equation}
Since $\max(p_2,1+p_1-p_2)$ and $p_3$ are both less than one, a random
geodesic again preferentially sees one dimension of space.

One might expect this behavior to be reflected in the heat kernel and the
spectral dimension.  The exact form of heat kernel for Kasner space is
not known, but Futamase \cite{Futamase} and Berkin \cite{Berkin}
have looked at different approximations.  Both find behavior of the form
\begin{equation}
K(x,x;s) \sim \frac{1}{4\pi s^2}\left[ 1 + \frac{a}{t^2}\,s + \dots \right] .
\label{Ca12}
\end{equation}
For a fixed time $t$, one can always find $s$ small enough that the first
term in (\ref{Ca12}) dominates: the heat kernel is a classical object, and 
the underlying classical spacetime is still four-dimensional.  For a fixed
return time $s$, on the other hand, one can always find a time $t$ small 
enough that the second term dominates, leading to an effective spectral 
dimension of two.\footnote{One might worry that at smaller $t$, even
higher powers of $s$ dominate.  But these lead to terms in the propagator
that go as positive powers of the geodesic distance, and are irrelevant 
for short distance singularities, light cone behavior, and the like.}  The
idea that the ``effective infrared dimension'' might differ from four goes
back to work by Hu and O'Connor \cite{Hu}, but the relevance to 
short-distance quantum gravity was not fully appreciated at that time.

One can also investigate this issue by using the Seeley-DeWitt expansion
of the heat kernel \cite{DeWitt2,Gibbons,Vassilevich},
\begin{equation}
K(x,x,s)  \sim \frac{1}{4\pi s^2}\left( [a_0] + [a_1]s + [a_2]s^2 
   +\dots \right) .
\label{Ca13}
\end{equation}
The ``Hamidew coefficient'' $[a_1]$ is proportional to the scalar 
curvature, and vanishes for an exact vacuum solution of the field 
equations.  In the presence of matter, however, the scalar curvature
will typically increase as an inverse power of $t$ as $t\rightarrow0$
\cite{BKL}; this growth is slow enough to not disrupt the BKL behavior
of the classical solutions near $t=0$, but it will nevertheless give a
diverging contribution to $[a_1]$.

The short-distance BKL behavior suggested by the Wheeler-DeWitt
equation may thus offer an explanation for the dimensional reduction 
of quantum gravity at the Planck scale.  The dynamics picks out an 
essentially random dominant spatial direction at each point, whose 
existence is reflected in the behavior of the heat kernel and the 
propagators.  To probe this picture further, though, we must better 
understand the underlying physics.

\section{Asymptotic silence?}

The BKL picture was originally developed in a very different context 
from the one I am considering here, as a study of the Universe near 
an initial spacelike singularity.  In that setting, the key physical 
ingredient is ``asymptotic silence'' \cite{HUR}, the strong focusing 
of light by the singularity that collapses light cones and shrinks 
particle horizons.  It is this behavior that decouples neighboring 
points and leads to the ultralocal form of the equations of motion.

The similar decoupling of neighboring points in the strongly coupled 
Wheeler-DeWitt equation suggests that a similar physical mechanism
may be at work.  To see whether this is the case will require much
deeper investigation.  As a first hint, though, consider the classical
Raychaudhuri equation,
\begin{equation}
\frac{d\theta}{d\lambda} = -\frac{1}{2}\theta^2 
      - \sigma_\alpha{}^\beta \sigma_\beta{}^\alpha
     + \omega_{\alpha\beta}\omega^{\alpha\beta} 
      - R_{\alpha\beta}k^\alpha k^\beta ,
\label{Ca14}
\end{equation}
for the expansion of a bundle of light rays.  We do not know the quantum
version of this relation, but if we naively treat (\ref{Ca14}) as an 
operator equation in the Heisenberg picture and take the expectation
value, we see that quantum fluctuations in the expansion and shear 
always focus geodesics:
\begin{equation}
\langle\theta^2\rangle = \langle\theta\rangle^2 + (\Delta\theta)^2 ,
\label{Ca15}
\end{equation}
with a similar equation for $\sigma$.  

How strong is this focusing?  Roughly speaking, the expansion $\theta$ 
is canonically conjugate to the cross-sectional area of the congruence 
\cite{Epp}: as the trace of an extrinsic curvature, it is conjugate to 
the corresponding volume element.  Keeping track of factors of $\hbar$ 
and $G$, one finds an uncertainty relation
\begin{equation}
\Delta{\bar\theta}\Delta A \sim \ell_p ,
\label{Ca16}
\end{equation}
where $\bar\theta$ is the expansion averaged over a Planck distance
along the congruence.  If, near the Planck scale, the area uncertainty 
is of order $\ell_p^2$---as one might expect from theories such as loop 
quantum gravity in which the area spectrum is quantized---this would 
imply fluctuations of $\theta$ of order $1/\ell_p$, giving strong focusing.

This argument is, of course, only suggestive.  In particular, I have 
ignored the need to renormalize products of operators such as 
$\theta^2$, and have neglected the effects of the twist and curvature 
terms in (\ref{Ca14}).  But while the result is not yet established, it 
is at least plausible that quantum fluctuations at the Planck scale, 
``spacetime foam,'' could lead to strong focusing of geodesics, and 
thus to short-distance asymptotic silence.
 
\section{A new picture}

If the proposals of the two preceding sections are correct---if 
spacetime foam strongly focuses geodesics at the Planck scale, leading 
to the BKL behavior predicted by the strongly coupled Wheeler-DeWitt 
equation---they suggest a novel picture of  small-scale spacetime.  At 
each point, the dynamics picks out a ``preferred'' spatial direction, 
leading to approximately (1+1)-dimensional local physics.  The 
preferred directions are presumably determined by initial conditions, 
but because of the chaotic behavior of BKL bounces, they are quickly 
randomized.  From point to point, these directions vary continuously, 
but oscillate rapidly \cite{Montani}.  Space at a fixed time is thus 
threaded by rapidly fluctuating lines, and spacetime by two-surfaces, 
and the leading behavior of the physics is described by an approximate 
``dimensional reduction'' to these surfaces.

There is a danger here, of course: the process I have described breaks
Lorentz invariance at the Planck scale, and even small violations at
such scales can have observable effects at larger scales \cite{Mattingly}.
Note, though, that the symmetry violations in the present scenario
vary rapidly and essentially stochastically in both space and time.
Such ``nonsystematic'' Lorentz violations are harder to study, but 
there is evidence that they lead to much weaker observational
constraints \cite{Basu}.

The scenario I have presented is still very speculative, but I believe 
it deserves further investigation.  One avenue might be to use results
from the eikonal approximation \cite{tHooft,Verlinde,Kabat}.  In
this approximation, developed to study very high energy scattering,
 a similar dimensional reduction takes place, with drastically disparate 
time scales in two pairs of dimensions.  Although the context is very
different, the technology developed for this approximation could 
prove useful for the study of Planck scale gravity.

\begin{theacknowledgments}
I would like to thank Beverly Berger, David Garfinkle and Bei-Lok Hu
for valuable advice at the beginning of this investigation.
This work was supported in part by U.S.\ Department of Energy grant
DE-FG02-91ER40674.
\end{theacknowledgments}


\begin{thebibliography}{99}

\bibitem{AJL}
J.~Ambj{\o}rn, J.~Jurkiewicz, and R.~Loll, \emph{Phys.\ Rev.\ Lett.} 
  \textbf{85}, 924 (2000). eprint hep-th/0002050.
\bibitem{AJL2}
J.~Ambj{\o}rn, J.~Jurkiewicz, and R.~Loll, \emph{Phys.\ Rev.\ Lett.} 
  \textbf{93}, 131301 (2004), eprint hep-th/0404156.
\bibitem{AJL3} 
J.~Ambj{\o}rn, J.~Jurkiewicz, and R.~Loll, \emph{Phys.\ Rev.}
  \textbf{D72} 064014 (2005), eprint hep-th/0505154.
\bibitem{Regge} 
   T.~Regge, \emph{Nuovo Cimento A} \textbf{19}, 558 (1961).
\bibitem{Williams} 
   M.~Rocek and R.~M.~Williams, \emph{Phys.\ Lett.} \textbf{B104},
   31 (1981).
\bibitem{Loll} 
   R.~Loll, \emph{Living Rev.\ Relativity} \textbf{1}, 13 (1998),
   eprint gr-qc/9805049.
\bibitem{Kommu} 
   R.~Kommu, in preparation.
\bibitem{AJL4} 
   J.~Ambj{\o}rn, J.~Jurkiewicz, and R.~Loll, \emph{Phys.\ Lett.}
   \textbf{B607}, 205 (2005), eprint hep-th/0411152.
\bibitem{spec} 
J.~Ambj{\o}rn, J.~Jurkiewicz, and R.~Loll, \emph{Phys.\ Rev.\ Lett.} 
  \textbf{95}, 171301 (2005), eprint hep-th/0505113.
\bibitem{Synge} 
   J.~L.~Synge, \emph{Relativity: The General Theory}, North-Holland, 
   Amsterdam, 1960.
\bibitem{Weinberg} 
   S.~Weinberg, ``Ultraviolet divergences in quantum theories of 
   gravitation,'' in \emph{General Relativity: An Einstein Centenary 
   Survey}, edited by S.~W.~Hawking and W.~Israel, Cambridge University 
   Press, Cambridge, 1979, pp.~790--831.
\bibitem{Reuter}
   M.~Reuter and F.~Saueressig, \emph{Phys.\ Rev.} \textbf{D65}, 
   065016 (2002), eprint hep-th/0110054.
\bibitem{Litim} 
   D.~F.~Litim, \emph{Phys.\ Rev.\ Lett.} \textbf{92}, 201301 (2004),
   eprint hep-th/0312114.
\bibitem{Nieder} 
   M.~Niedermaier, \emph{Class.\ Quant.\ Grav.} \textbf{24}, R171 (2007),
   eprint gr-qc/0610018.
\bibitem{Reuter2} 
   O.~Lauscher and M.~Reuter, \emph{JHEP} \textbf{0510} 050 (2005),
   eprint hep-th/0508202.
\bibitem{Modesto}
   L.~Modesto, `` Fractal Structure of Loop Quantum Gravity,'' eprint
   arXiv:0812.2214 [gr-qc].
\bibitem{Atick} 
   J.~J.~Atick and E.~Witten, \emph{Nucl.\ Phys.} \textbf{B310}, 291 
   (1988).
\bibitem{Horava}
   P.~Ho{\v r}ava, \emph{Phys.\ Rev.} \textbf{D79}, 084008 (2009),
   eprint arXiv:0901.3775 [hep-th].
\bibitem{Horava2}
   P.~Ho{\v r}ava, \emph{Phys.\ Rev.\ Lett.} \textbf{102}, 161301 (2009),
   eprint arXiv:0902.3657 [hep-th].
\bibitem{Isham} 
   C.~J.~Isham, \emph{Proc.\ R.\ Soc.\ London} \textbf{A351}, 209 (1976).
\bibitem{DeWitt}
   B.~S.~DeWitt, \emph{Phys.\ Rev.} \textbf{160}, 1113 (1967).
\bibitem{Pilati} 
   M.~Pilati, ``Strong Coupling Quantum Gravity: An Introduction,'' in 
   \emph{Quantum structure of space and time}, edited by M.~J.~Duff and 
   C.~J.~Isham, Cambridge University Press, Cambridge, 1982, pp.~53--69.
\bibitem{Maeda} 
   K.~Maeda and M.~Sakamoto, \emph{Phys.\ Rev.} \textbf{D54}, 1500 
   (1996), eprint hep-th/9604150.
\bibitem{HPT} 
   M.~Henneaux, M.~Pilati, and C.~Teitelboim, \emph{Phys.\ Lett.} 
   \textbf{110B}, 123 (1982).
\bibitem{Teitelboim} 
   C.~Teitelboim, \emph{Phys.\ Rev.} \textbf{D25}, 3159 (1982).
\bibitem{Pilati2} 
   M.~Pilati, \emph{Phys.\ Rev.} \textbf{D26}, 2645 (1982); 
   \emph{Phys.\ Rev.} \textbf{D28}, 729 (1983).
\bibitem{Rovelli}
   C.~Rovelli, \emph{Phys.\ Rev.} \textbf{D35}, 2987 (1987).
\bibitem{Husain} 
   V.~Husain, \emph{Class.\ Quant.\ Grav.} \textbf{5}, 575 (1988).
\bibitem{Francisco} 
   G.~Francisco and M.~Pilati, \emph{Phys.\ Rev.} \textbf{D31}, 241 
   (1985).
\bibitem{Salopek} 
   D.~S.~Salopek, \emph{Class.\ Quant.\ Grav.} \textbf{15}, 1185 (1998),
   eprint gr-qc/9802025.
\bibitem{Kirillov} 
   A.~A.~Kirillov, \emph{Int.\ J.\ Mod.\ Phys} \textbf{D3}, 431 (1994).
\bibitem{Henneaux} M.~Henneaux, \emph{Bull.\ Math.\ Soc.\ Belg.} 
   \textbf{31}, 47 (1979).
\bibitem{Helfer} A.~Helfer et al., \emph{Gen.\ Rel.\ Grav.} \textbf{20},
   875 (1988).
\bibitem{BKL}
   V.~A.~Belinskii, I.~M.~Khalatnikov, and E.~M.~Lifshitz, 
   \emph{Adv.\ Phys.} \textbf{19}, 525 (1970); \emph{Adv.\ Phys.}
   \textbf{31}, 639 (1982).
\bibitem{HUR} 
   J.~M.~Heinzle, C.~Uggla, and N.~R{\"o}hr, \emph{Adv.\ Theor.\ Math.\ 
   Phys.} \textbf{13}, 293 (2009), eprint gr-qc/0702141.
\bibitem{Misner}
   C.~W.~Misner, \emph{Phys.\ Rev.\ Lett.} \textbf{22}, 1071 (1969).
\bibitem{Chernoff} 
   D.~F.~Chernoff and J.~D.~Barrow, \emph{Phys.\ Rev.\ Lett.} \textbf{50}, 
   134 (1983).
\bibitem{Kirillov2} 
   A.~A.~Kirillov and G.~Montani, \emph{Phys.\ Rev.} \textbf{D56}, 
   6225 (1997).
\bibitem{Futamase} 
   T.~Futamase,  \emph{Phys.\ Rev.} \textbf{D29}, 2783 (1984).
\bibitem{Berkin} 
   A.~L.~Berkin, \emph{Phys.\ Rev.} \textbf{D46}, 1551 (1992).
\bibitem{Hu} 
   B.~L.~Hu and D.~J.~O'Connor,  \emph{Phys.\ Rev.} \textbf{D34}, 2535 
   (1986).
\bibitem{DeWitt2} 
   B.~S.~DeWitt, `` Dynamical theory of groups and fields,'' in 
   \emph{Relativity, Groups and Topology}, edited by C.~DeWitt and
   B.~DeWitt, Gordon and Breach, New York, 1964, pp.~587--820.
\bibitem{Gibbons} 
   G.~W.~Gibbons, ``Quantum field theory in curved spacetime,'' in 
   \emph{General Relativity: An Einstein Centenary Survey}, edited by 
   S.~W.~Hawking and W.~Israel, Cambridge University Press, Cambridge, 
   1979, pp.~639--79.
\bibitem{Vassilevich}
   D.~V.~Vassilevich, \emph{Phys.\ Rept.} \textbf{388}, 279 (2003),
   eprint hep-th/0306138.
\bibitem{Epp} 
   R.~Epp, ``The Symplectic Structure of General Relativity in the 
   Double Null (2+2) Formalism,'' eprint gr-qc/9511060.
\bibitem{Montani}
   G.~Montani, \emph{Class.\ Quant.\ Grav.} \textbf{12}, 2505 (1990).
\bibitem{Mattingly}
   D.~Mattingly, \emph{Living Rev. \ Relativity} \textbf{8},  5 (2005),
   eprint gr-qc/0502097.
\bibitem{Basu}
   S.~Basu and D.~Mattingly, \emph{Class.\ Quant.\ Grav.} \textbf{22},
   3029 (2005), eprint astro-ph/0501425.
\bibitem{tHooft}
   G.~'t Hooft, \emph{Phys. \ Lett.} \textbf{ 198B}, 61 (1987).
\bibitem{Verlinde}
   H.~L.~Verlinde and E.~P.~Verlinde,  \emph{Nucl.\ Phys.} \textbf{B371},
   246 (1992), eprint hep-th/9110017.
\bibitem{Kabat}
   D.~Kabat and M.~Ortiz, \emph{Nucl.\ Phys.} \textbf{B388}, 570 (1992),
   eprint hep-th/9203082.
\end{thebibliography}
\end{document}